\documentstyle[12pt]{article}

\newcommand{\be}{\begin{equation}}
\newcommand{\ee}{\end{equation}}
\newcommand{\ba}{\begin{eqnarray}}
\newcommand{\ea}{\end{eqnarray}}
\begin{document}
\def\pct#1{\input epsf \centerline{ \epsfbox{#1.eps}}}

\begin{titlepage}
\hbox{\hskip 12cm ROM2F-96/23  \hfil}
\hbox{\hskip 12cm \today \hfil}
\vskip 1.5cm
\begin{center} 
{\Large  \bf  Open \ Descendants \ in \ Conformal \ Field \ Theory}

\vspace{1.8cm}
 
{\large \large  Augusto Sagnotti \ \ and \ \ Yassen S. Stanev
\footnote{I.N.F.N.  Fellow,
on Leave from Institute for Nuclear Research and Nuclear Energy, Bulgarian 
Academy of Sciences, BG-1784 Sofia, BULGARIA.}}

\vspace{0.8cm}

{\sl Dipartimento di Fisica\\
Universit{\`a} di Roma \ ``Tor Vergata'' \\
I.N.F.N.\ - \ Sezione di Roma \ ``Tor Vergata'' \\
Via della Ricerca Scientifica, 1 \ \
00133 \ Roma \ \ ITALY}
\vspace{0.4cm}
\end{center}
\vskip 0.6cm
\abstract{ 
Open descendants extend
Conformal Field Theory to unoriented surfaces with boundaries.
The construction rests on two types of generalizations of the fusion algebra.  
The first is needed even in the relatively simple case of diagonal models. It leads to a
new tensor that satisfies the fusion algebra, but whose entries are signed integers. The
second is needed when dealing with non-diagonal models, where Cardy's ansatz does
not apply. It leads to a new tensor with positive integer entries, that
satisfies a set of polynomial equations and encodes the classification
of the allowed boundary operators.}

\vskip 1.4cm
\begin{center}
Based on Talks Presented by A. Sagnotti at the CERN Meeting on STU
Dualities, Dec. 1995, and by Ya.S. Stanev at the 1996 Rome
Triangle Meeting, March 1996.
\end{center}
 \vfill
\end{titlepage}
\makeatletter
\@addtoreset{equation}{section}
\makeatother
\renewcommand{\theequation}{\thesection.\arabic{equation}}
\addtolength{\baselineskip}{0.3\baselineskip} 

\vskip 24pt
\section{Introduction}

A byproduct of String Duality is a renewed interest in 
open-string theories \cite{pol}. These \cite{open} provide also a
relatively  simple instance of the
interplay between different dimensions, an emerging aspect of Field
Theory that is rapidly affecting long-held views on such fundamental issues as chirality
and anomalies
\cite{diffdim}. We have decided to depart slightly from the CERN
talk, a general survey on open-string models and Chan-Paton charges \cite{cp}, 
to confine our attention to the open descendants of Rational Conformal
Field Theory \cite{cargese,bs,bps,fps,pss,pss3}. These extend
Rational Conformal Field Theory \cite{ms} to surfaces with
boundaries \cite{cardy,cardylew,lew} and/or crosscaps, and exhibit neatly many
interesting features. The resulting presentation should complement
refs. \cite{reviews}. The following four sections, devoted to the four 
amplitudes that define the spectrum, include two instructive examples drawn from our
previous study \cite{pss} of the $ADE$ series \cite{ciz} of $SU(2)$ WZW models
\cite{wzw}. In Section 2 we
review the  structure of the oriented closed sector. This involves the torus amplitude,
by now standard knowledge, but sets the stage for the construction of
open descendants and allows us to introduce our notation and conventions.  
In Section 3 we discuss
the Klein-bottle projection of the closed spectrum. In Section 4 we discuss
the annulus amplitude and the related issue of boundary operators,
including our recent work on boundary operators in non-diagonal models \cite{pss3}.  
This part
should also be of some interest in Condensed Matter Physics, where boundary operators
can describe the effects of impurities near criticality \cite{aflud}.  Finally, in Section
5 we discuss the  M\"obius-strip projection of the open spectrum.

\vskip 24pt
\section{The Torus Amplitude}

\vskip 15pt
\input epsf \centerline{ \epsfbox{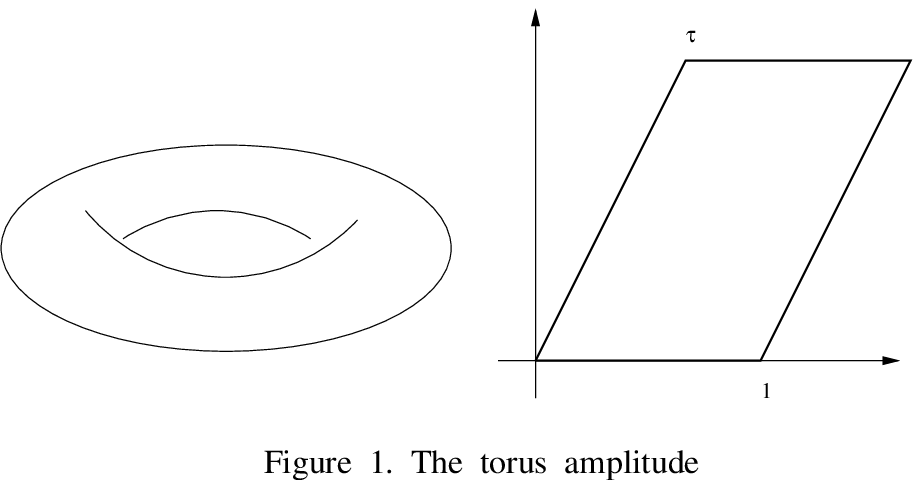}}
\vskip 15pt

The first amplitude of our construction is
the torus of fig. 1. Its fundamental polygon, a parallelogram, has
one vertex at the origin and two sides ending at $1$ and $\tau$, with $Im(\tau) >0$.
We restrict our attention to a two-dimensional
rational conformal field theory with a symmetry algebra 
${\cal A} \times {\bar{\cal A}}$, such that

- the left and right symmetry algebras are identical;

- the symmetry algebra is maximally extended.

\noindent
The two assumptions imply a one-to-one correspondence between the
two-dimensional field content and the chiral content, described by a
finite  set of primary fields
 $\{ \varphi_i \}$ of ${\cal A}$. These 
can be neatly represented by the (Virasoro reduced) characters of 
the symmetry algebra
\be
\chi_i (\tau) = Tr_{{\cal H}_i} e^{2 \pi i \tau \hat{L}_0} \quad ,
\label{character}
\ee
where the trace is over the corresponding ${\cal A}$ module.
Since the left and right symmetry algebras coincide, 
the same is true for the corresponding
sets of characters, as well as for the primary fields.  
The left-right symmetry is a necessary condition to construct 
open descendants. On the other hand, the assumption of maximality can  
well be relaxed, although it simplifies 
considerably the resulting formulas. For instance, the $D_{even}$ 
models may be treated as $su(2)$ models or as extended symmetry ones.  

The two generators of the $PSL(2,Z)$ group of modular transformations 
act linearly on the characters, according to 
\be
\chi_i \left(- \ {1 \over \tau}\right) \ = \ \sum_j \ S_{ij} \ \chi_j(\tau) \quad
\qquad
\chi_i (\tau +1) \ = \ \sum_j \ T_{ij} \ \chi_j(\tau)  \quad ,
\label{modular}
\ee
where $S$ and $T$ are unitary symmetric matrices that satisfy
$S^2 = (ST)^3 = C$, with $C$ the conjugation matrix and $C^2=1$. 
All the infinitely many choices of fundamental cell 
in the lattice generated by the parallelogram in fig. 1 may be related by the two
transformations of eq. (\ref{modular}). Modular invariance thus makes the
choice of fundamental cell immaterial.  
The matrix $T$, diagonal in the basis of
characters, encodes the conformal weights of the primary fields.

The two-dimensional field content can be read from the modular invariant
partition function 
\be
{Z_T} \ = \ \sum_{i,j} \ \chi_i \ X_{ij} \ \chi_j \quad \qquad  S X S^{\dagger} = X
\quad \qquad  T X T^{\dagger} = X \quad , \label{torus}
\ee
where the non-negative integers $X_{ij}$ count the multiplicities of the 
two-dimensional fields $\varphi_{i,\bar j} (z,\bar z)$ in the bulk spectrum.
The assumption that the symmetry algebra $\cal A$ be maximal
implies that the torus partition function, if not diagonal, is at most a
permutation invariant.  Thus, we are effectively restricting our attention to 
models with
\be
X_{ij} = \delta_{i \sigma (j)} \quad , \label{permutation}
\ee
where $\sigma (j)$ denotes a permutation of the labels $j$. Each chiral
character is always coupled to a single antichiral one, and there are no
multiplicities larger than one in the bulk spectrum. 

Denoting by $[\varphi]$ the conformal family of $\varphi$, one has the fusion
algebra
\be
[\varphi_i] \times [\varphi_j] \ = \ \sum_k \ {N_{ij}}^k \ [\varphi_k] \quad ,
\label{fusion}
\ee
where the Verlinde formula \cite{verlinde}
\be
{N_{ij}}^k \ = \ \sum_{\ell} \ { S_{i \ell} S_{j \ell} S_{k \ell}^{\dagger} 
\over S_{1 \ell}} \label{verlinde}
\ee
relates the fusion-rule coefficients ${N_{ij}}^k$ to the modular matrix $S$. The
integers ${N_{ij}}^k$ encode the basic information
in the OPE \cite{bpz} of any pair of fields.

It should be appreciated that modular invariance  
plays also the role of a completeness condition for two-dimensional fields.
Any set of fields  whose  
torus amplitude is not modular invariant, even if closed under the 
operator product, should 
be consistently enlarged by the inclusion of other fields.
In Section 4 we shall see that extending the completeness requirement 
to the open
sector reduces the classification problem of the allowed boundary states to
a set of polynomial equations closely related to the Verlinde algebra \cite{pss3}. 

We shall illustrate the construction for the $k=2$ ($A_3$) 
and for the $k=6$ ($D_5$)  $su(2)$ WZW models, the simplest ones in the $A$ 
and $D_{odd}$ series. General formulae valid for the whole series,
as well as the treatment of $D_{even}$ and $E$ models, may be found in 
refs. \cite{pss,pss3}.
We label the characters by shifted $su(2)$ weights, related to the isospin 
by $i=2 I +1$. In this notation the corresponding modular invariants are:

- for the $A_3$ model
\be
Z_T^{A_3} = |\chi_1|^2 + |\chi_2|^2 + |\chi_3|^2 \quad ;
\label{Ta3}
\ee

- for the $D_5$ model
\be
Z_T^{D_5} = |\chi_1|^2 + |\chi_3|^2 + |\chi_5|^2 + |\chi_7|^2 + |\chi_4|^2
+\chi_2 {\bar \chi_6} + \chi_6 {\bar \chi_2} \quad .
\label{Td5}
\ee

\vskip 24pt
\section{The Klein-Bottle Amplitude}

\vskip 15pt
\input epsf \centerline{ \epsfbox{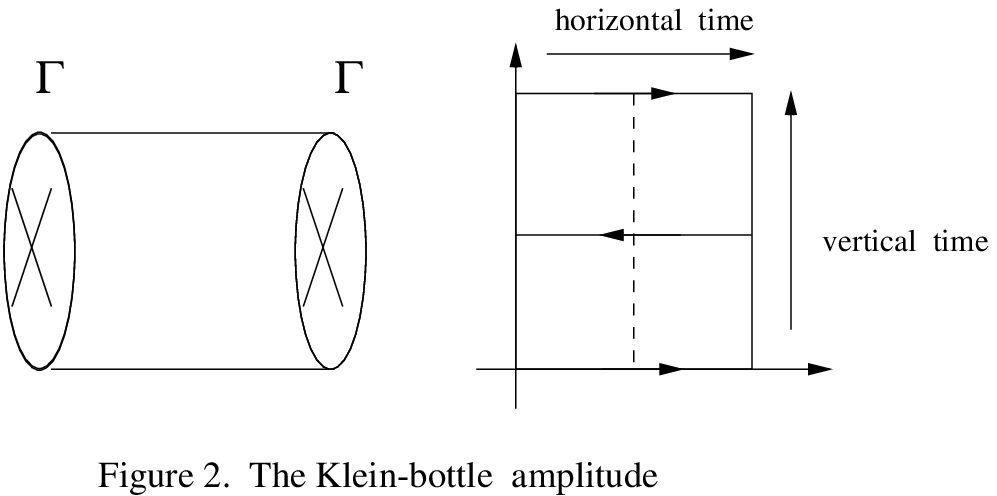}}
\vskip 15pt

The construction of open descendants involves 
surfaces with boundaries and crosscaps. The crosscap, the simplest 
non-orientable surface, may be pictured as a disk with diametrically 
opposite points identified.
The Klein bottle is a closed non-orientable surface with
two crosscaps and no handles. 

The introduction of crosscaps breaks the ${\cal A} \times
{\bar{\cal A}}$ symmetry of the 2-dimensional model. 
Here we confine our attention
to  crosscaps preserved by a residual ${\cal A}$ symmetry, and in Section 4 we
shall extend this requirement to the boundaries. There
are, however, two crucial differences between the crosscap and the more
familiar case of a boundary : the involution that defines the crosscap has no fixed
points, and the crosscap itself may not be localized. 

The crosscap interchanges chiral and 
antichiral parts of two-dimensional fields, and thus defines an involution 
\be
\Omega : \varphi_{i \bar i} \rightarrow \epsilon_{i} \
\varphi_{\bar i i} \quad , \quad \qquad  \epsilon_{i} = \epsilon_{\bar i} = \pm 1
\quad .
\label{ccaction}
\ee
Since we have
a one-to-one correspondence between two-dimensional and chiral spectra, the
action of this 
involution can be extended to the chiral characters. 
Consistency with the fusion rules demands that the signs $\epsilon_i$
satisfy the constraints
\be
{\rm if} \quad N_{i j k } \ne 0  \qquad {\rm then } \qquad 
\epsilon_{i} \epsilon_{j} \epsilon_{k} \ = \ 1 \quad . \label{signs}
\ee

The Klein bottle is pictured in figure 2, together with the
corresponding polygon, that in this case is a rectangle, so that the Teichm\"uller
parameter $\tau$ is purely imaginary. The two vertical sides are
directly identified, while the two horizontal ones are identified after
an orientation reversal.
There are two canonical inequivalent choices of time, and consequently
the ultraviolet point $\tau = 0$ is {\it not} excluded form the moduli space.
Referring to fig. 2, ``vertical'' time exhibits the propagation of closed strings
undergoing a turnover, and one has 
\be
K \ = \  \sum_i \ \chi_i \ K^i \quad . \label{kleind}
\ee
The $K^i$, integers constrained to satisfy $| K^i | = X_{i i}$, 
are always $0$ or $ \pm 1$ for a permutation invariant. They are 
related to the $\epsilon_{i}$, and should also satisfy eqs. (\ref{signs}).

A modular $S$ transformation to ``horizontal'' time turns eq. (\ref{kleind}) into
the transverse (vacuum)
channel that describes the propagation of closed strings on a tube terminating
at the two crosscaps,
\be
\tilde K \ = \  \sum_i \ \chi_i \ {\Gamma_i}^2 \quad . \label{kleint}
\ee
The reflection coefficients
$\Gamma_i$ determine the one-point functions of the bulk fields in front
of the  crosscap. 

In fig. 2 the two crosscaps may be associated with the left border of the
rectangle and with the dashed line. The transverse channel may then
be identified with the
portion of the double cover limited by these two lines.

The  n-point functions of the two-dimensional fields in 
front of a crosscap can be expressed in terms of the corresponding 
2n-point chiral 
conformal blocks. In particular, the $\Gamma_i$
can be non-vanishing only for fields whose left and right conformal 
dimensions coincide, $( \Delta_i = \Delta_{\bar i} )$, and only if $\epsilon_{i} = 1$.
The consistency of two-point functions with the involution of eq. (\ref{ccaction}) 
implies the ``crosscap constraint'' \cite{fps,pss}, a set of 
linear relations for the $\Gamma_i$ and the signs 
$\epsilon_{j}$ involving also the two-dimensional structure constants and the 
fusion matrix $F$. In general, these determine completely the $\Gamma_i$, and
thus, after an $S$ modular transformation, the direct-channel coefficients 
$K^i$.
Although it is usually simpler to solve directly the system of eqs.
(\ref{signs}) for $K^i$, this does not determine the signs of the 
$\Gamma_i$ that, as we shall see, are needed in the M\"obius-strip amplitude.

In the projected spectrum, ${1 \over 2} (Z_T + K)$, the 
Verma modules are (anti)symmetrized under left-right interchange.
 Multiple solutions, whenever present, describe inequivalent  projections
of the closed spectrum, and the complete open descendants have
correspondingly different spectra.  

There are two possible choices for the Klein bottle partition functions 
both in the $A_3$ and in the $D_5$ models. Anticipating the structure 
of the resulting Chan-Paton groups, we shall denote them by (r)eal
and (c)omplex respectively. The direct channel expressions corresponding 
to the modular invariants of eqs. (\ref{Ta3}) and (\ref{Td5}) are then
\ba
K_r^{A_3} \ &=& \  \chi_1 - \chi_2 + \chi_3 \quad,\\
K_c^{A_3} \ &=& \  \chi_1 + \chi_2 + \chi_3 \quad ,\\
K_r^{D_5} \ &=& \  \chi_1 + \chi_3 + \chi_5 + \chi_7 - \chi_4 \quad,\\
K_c^{D_5} \ &=& \  \chi_1 + \chi_3 + \chi_5 + \chi_7 + \chi_4 \quad .
\ea
It is instructive to compare these expressions with the signs $\epsilon_i$
of eq. (\ref{ccaction}), that in both models are 

- $\epsilon_i = 1$ for all $i$ in the real case;

- $\epsilon_i = (-1)^{i-1}$  in the complex case.

The relative factor $(-1)^{2I}$ between $K_i$ and $\epsilon_i$ is 
introduced by the $su(2)$ structure of the fields. It should be
appreciated that, differently from the $K_i$, the $\epsilon_i$ are 
well defined for all fields.

\vskip 24pt
\section{The Annulus Amplitude}

\vskip 15pt
\input epsf \centerline{ \epsfbox{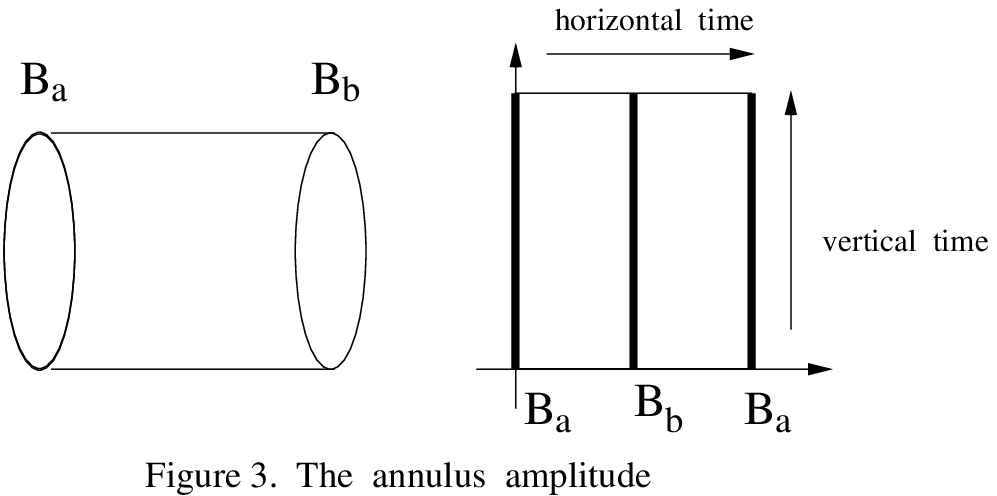}}
\vskip 15pt

Let us now turn to the annulus amplitude of fig. 3, whose
fundamental polygon is again a rectangle. The horizontal sides
identified, while the vertical sides represent the two boundaries.
The main issue of this Section is to indicate how to construct, 
for each given model, the allowed types of boundary states.

The annulus is doubly covered by a torus whose 
fundamental polygon is also a rectangle, and therefore there is again
only one real Teichm\"uller parameter.  The two canonical
choices of ``time'', that referring to fig. 3 may be termed
``horizontal'' and ``vertical'', have quite distinct roles, and consequently the
range of the parameter is again the whole positive imaginary axis.  As in the
Klein-bottle amplitude, the
ultraviolet point $\tau=0$ is {\it not} excluded from the modular region. The two
choices of time have quite distinct roles: vertical time exhibits a vacuum amplitude
for open strings, while horizontal time exhibits an amplitude for closed strings
propagating between two boundaries.  Therefore, this diagram gives
information on the boundary operators, a new sector of the operator space, but at the
same time it rests on the known spectrum of the closed model.  This crucial property,
discovered very early in the bosonic string, played a central
role in the proposal of ref. \cite{cargese}.
Indeed, the annulus amplitude determines to a large extent the open sector of the theory in
terms of its closed sector. The other two surfaces, the Klein bottle and the
M\"obius strip, lead to additional, related truncations of 
closed and open  spectra. 

The introduction of boundaries breaks the total
${\cal A} \times {\bar{\cal A}}$ symmetry of the two-dimensional model, and
can actually do so to different extents, leading to different classes
of descendants. The restriction to conformally invariant
boundaries is natural in String Theory. As in Section 3, we
actually confine our attention to the maximal case, where the residual symmetry
is ${\cal A}$. Only
one Virasoro algebra acts on the  fields in this case, with the same central charge as
both the chiral  and antichiral ones. Hence, both the representations and the
characters  of the bulk and boundary algebras coincide.
This implies that the n-point functions of two-dimensional fields 
solve the same differential equations as the 2n-point chiral 
conformal blocks, while the corresponding partition function is 
linear (rather than sesquilinear) in the same characters.

Following Cardy \cite{cardy}, one can introduce boundary conditions
corresponding to different boundary states of the residual 
${\cal A}$ symmetry, as well as boundary operators $\psi_i^{a b}$ that 
mediate between 
them. On the annulus one can define two
different partition functions. The
direct channel (corresponding to the ``vertical'' time in fig. 3) exhibits
the propagation of open strings with ends on the two boundaries, and
\be
A \ = \  \sum_{i,a,b} \ \chi_i \ A_{a b}^i \  n^a \ n^b \quad ,
\label{annulusd}
\ee 
where the non-negative integers $A_{a b}^i$ count the multiplicities
of the boundary operators $\psi_i^{a b}$ in the spectrum. We
have allowed for multiplicities $n^a$ associated with the boundaries, and the 
boundary operators can thus  be matrix valued.
In open-string models this extension
results in the introduction of Chan-Paton groups, restricted by the
factorization of
tree-level amplitudes to the three infinite families $U(n)$,
$O(n)$ and $USp(2n)$ \cite{cp}. The boundaries are valued in the fundamental
representations of these classical Lie groups, and one is to distinguish 
two cases. When
the group is $U(n)$, there are two inequivalent choices for the fundamental
representation, that can thus be associated with oriented boundaries. 
On the other hand, no orientation is  needed for $USp(2n)$ and
$O(2n)$, whose fundamental representations are (pseudo)real.
Alternatively, in applications to Statistical Mechanics, one may regard 
eq. (\ref{annulusd}) as a formal polynomial in the variables $n^a$, a generating
function for the  multiplicities of the allowed boundary fields.
 
The transverse or vacuum channel (corresponding to the ``horizontal'' time
in fig. 3) has a very different interpretation: it exhibits
the propagation of closed strings along a tube terminating at the two
boundaries.  Since the closed spectrum undergoes a pair of reflections, this
amplitude is linear in the characters, with coefficients that are perfect squares
\be
\tilde A = \sum_{i} \ \chi^i \ {\left[ \sum_a \ B_{i a} \ n^a \right]}^2
\quad . 
\label{annulust}
\ee 
$B_{i a}$, the reflection coefficient for the sector $i$ of the bulk
spectrum from a boundary of type $a$, is proportional to the corresponding 
disk one-point function. The
sum within brackets is the total reflection coefficient for the sector $i$. 

Since $\tilde A$ is related to $A$ by a modular $S$
transformation, the open spectrum is closely linked to the disk one-point functions.
Cardy's ansatz \cite{cardy} determines the open spectrum when the  matrix $X$ in
eq. (\ref{torus}) is the conjugation matrix $C$. In this case both the open
sectors and the types of boundaries are in one-to-one correspondence with the
bulk spectrum, $ A_{i j}^k = N_{ij}^k$, and thus the fusion-rule coefficients
count the available boundary operators.

In the presence of boundaries, one can introduce two new types of operator products
\cite{cardylew,lew}. The
first,
\be
\psi_i^{a b} \ \psi_j^{b c} \ = \ \sum_{\ell} \ C_{i j \ell}^{a b c} \
\psi_{\ell}^{a c} \quad , \label{opebound}
\ee 
involves boundary fields only
and plays a central role in the definition of their disk amplitudes.  The second,
\be
\left.{ \varphi_{i \bar i} } \right|_a \ = \ \sum_{\ell} \ C_{(i \bar i) \ell}^{a} 
\ \psi_{\ell}^{a a} \quad , \label{opebulkto}
\ee 
involves both bulk and boundary fields, and
describes the behavior of bulk fields in front of the different boundaries.
The $B_{i a}$ in eq. (\ref{annulust}) are proportional 
to the $C_{(i \bar i) 1}^{a}$, with
coefficients that depend on the normalizations of the bulk and boundary fields.
The new structure constants, $C_{i j \ell}^{a b c}$ and $C_{(i \bar i) \ell}^{a}$,
and thus the reflection coefficients in $\tilde A$,
can all be determined by imposing the consistency of the full set of OPE's.
This leads to new sewing constraints, introduced in \cite{lew}. 
Some errors 
in the final expressions, related to the analytic continuations involved,
were corrected in \cite{pss3}, using also ref. \cite{rst}. 
These constraints extend the original ones of refs.
\cite{bpz,son} (the non-planar duality of bulk four-point amplitudes and the
modular invariance of torus one-point amplitudes), and lead to (quadratic or cubic)
relations for the structure constants. The method is quite powerful and gives very
detailed information on the theory, but it requires explicit expressions for the
braid and fusion matrices of the chiral conformal blocks.  These, however, are  known
only in a limited number of cases. This situation can be compared to a similar one
for the oriented closed sector. Modular invariant partition functions 
are classified for vast classes of models (see e.g. \cite{gannon} and references
therein), while the
two-dimensional structure constants are completely known only in a few cases
(essentially for some abelian models and for minimal and
$SU(2)$ WZW models).

This suggests an approach to the problem of classifying 
boundary states similar, in some respects, to the one followed for closed orientable
models.  Indeed, the integer
coefficients $A_{a b}^i$ in the direct-channel annulus amplitude of
eq. (\ref{annulusd}) have a nice interpretation as multiplicities of the 
corresponding boundary operators. Moreover, our two assumptions
on the structure of the symmetry algebra imply that they also count the different 
couplings $<a| \varphi_{i \bar i} |b>$,
since there is a one-to-one correspondence between two-dimensional and chiral
fields. For a complete set of boundary states
(we have already stressed that modular invariance plays a similar role for the
closed sector), one can derive the following set of
polynomial equations \cite{pss3} involving also  the fusion-rule coefficients  $N_{ij}^k$  
\ba
\sum_b \ {A^i}_{a}^{b} \ A_{b c}^j \ &=& \
\sum_k N^{ij}_k \  A_{a c}^k   \quad , \label{rel1} \\        
\sum_i \ A_{i a b} \ A_{c d}^i \ &=& \
\sum_i \ A_{i a c} \ A_{b d}^i \quad . \label{rel2}
\ea
Upper
and lower boundary indices are to be  distinguished whenever complex charges 
(corresponding to oriented boundaries) are present. Moreover, the matrix
$(A_1)_{a b} = (A_1)^{a b}$ is the metric for the boundary indices, since
it follows from eq. (\ref{rel1}) that  
$\sum_b A_{i a b} \ {A_1}^{b c} = {A_i}_a^c$, while $(A_1)_a^b = \delta_a^b $.

Eqs. (\ref{rel1}) and (\ref{rel2}) do not determine completely the 
matrices
$A_i^{a b}$, since they contain only chiral data.  The torus modular invariant of eq.
(\ref{torus}) is another crucial ingredient. Indeed, if for some
$j$ the matrix element $X_{j{\bar j}}$ of the modular invariant 
 vanishes, the tube can not support the corresponding mode, and the
bulk-to-boundary structure constants $C_{(j \bar j) 1}^{a}$  (and thus 
the reflection coefficients $B_{j a}$)
also vanish for all values of $a$. 
After a modular $S$ transformation this implies
that,  for all $a$ and $b$,
\be
\sum_i \ A_{i a b} \  S^{i}_j \ = \ 0 \quad . \label{rel3}
\ee
Hence, the $A_{i a b}$ are in general
linearly dependent matrices, while the number of different 
boundary states can be expressed rather neatly in terms of the 
corresponding modular invariant as $Tr(XC)$.
Eqs. (\ref{rel1})-(\ref{rel3}) have in general multiple solutions, but 
in all cases that we have analyzed explicitly they 
determine  $A_{i a b}$ up to the orientation of (pairs of) boundaries.
The solution corresponding to a given  $(A_1)_{a b}$ is unique, while different
choices for $(A_1)_{a b}$ are related by the action of a simple current that squares 
to the identity.

As an illustration, we shall again resort to the $A_3$ and 
$D_5$ $su(2)$ WZW models. Since the 
$A_3$ model is diagonal, in the real case
$A_{ijk} =  N_{ijk}$, and 
\be
A_r^{A_3} \ = \  \chi_1 (n_1^2 + n_2^2 + n_3^2) +
\chi_2 (2 n_1 n_2 + 2 n_2 n_3) +
\chi_3 (2 n_1 n_3 + n_2^2)  \quad .
\ee 
The second solution may be obtained acting on this spectrum 
with the simple current that corresponds
to $\chi_3$. It has a pair of complex charges that we denote by 
$n$ and $\bar n$, and  
\be
A_c^{A_3} \ = \  \chi_1 ( n_2^2 + 2 n {\bar n} ) +
\chi_2 (2 n_2 n + 2 n_2 {\bar n}) +
\chi_3 (n^2 + {\bar n }^2 + n_2^2) \quad .
\ee 
The equality $ \bar n = n $ ensures the positivity 
of the transverse-channel amplitude.

In the $D_5$ case the expressions are more complicated, and therefore 
a few remarks are in order. Since $\chi_2$ and $\chi_6$
enter off-diagonally the torus modular invariant (\ref{Td5}), eqs. 
(\ref{rel3}) limit to five the possible types of charges. 
Hence, the algebra of the boundary operators is 
different from the (chiral) bulk algebra, that involves seven types
of primary fields. In particular, there are boundary 
operators with multiplicities larger than one (some of the terms involving 
the $n_5$ charge below). As discussed in ref. \cite{pss3}, this corresponds
to a new phenomenon, whereby the boundary algebra is extended by
a simple current, in this case of dimension $3 \over 2$, that can not
extend the bulk algebra. 
The real assignment corresponding  to $(A_1)_{a b} = \delta_{a b}$ is 
\ba
\lefteqn{A_r^{D_5} \ =  \  
\chi_1 ( n_1^2 + n_2^2 + n_3^2 + n_4^2 + n_5^2 ) \ + \
(\chi_2 + \chi_6 )( 2 n_1 n_2  + 2 n_1 n_5 + 2 n_3 n_5 + 2 n_4 n_5) \ +}
\nonumber
\\ & & \chi_3 ( n_1^2 + 2 n_1 n_3 + 2 n_1 n_4 + 2 n_3 n_4 + 2 n_2 n_5 +
 2 n_5^2)
 \ + \nonumber \\
& & \chi_4 ( 4 n_1 n_5 + 2 n_2 n_3 + 2 n_3 n_5 + 2 n_2 n_4 + 2 n_4 n_5 )  \ +
\label{ard5} \\  
& &\chi_5 (n_1^2  + n_3^2  + n_4^2 + 2 n_5^2 + 2 n_1 n_3 + 2 n_1 n_4 +
 2 n_2 n_5 ) \ +
\
\chi_7 ( n_1^2 + n_2^2 + n_5^2 + 2 n_3 n_4 )  \nonumber \quad . 
\ea
The complex assignment has only one complex pair of charges ${\bar n} = n$
(replacing the $n_3$ and $n_4$ charges), and is obtained from the real one
acting with the simple current in $\chi_7$
\ba
\lefteqn{A_c^{D_5} \ =  \  
\chi_1 ( n_1^2 + n_2^2 + 2 n {\bar n}+ n_5^2 ) \ + \
(\chi_2 + \chi_6 )( 2 n_1 n_2  + 2 n_1 n_5 + 2 n n_5 + 2{\bar n} n_5) \ +}
\nonumber
\\ & & \chi_3 ( n_1^2 + n^2 + {\bar n}^2 + 2 n_1 n + 2 n_1 {\bar n}  + 2 n_2 n_5 +
 2 n_5^2)
 \ + \nonumber \\
& & \chi_4 ( 4 n_1 n_5 + 2 n_2 n + 2 n_2 {\bar n} + 2 n n_5 + 2 {\bar n} n_5 )  \ +
\label{acd5} \\  
& &\chi_5 (n_1^2 + 2 n_5^2 + 2 n_1 n + 2 n_1 {\bar n} +
 2 n_2 n_5 + 2 n {\bar n}) \ +
\
\chi_7 ( n_1^2 + n_2^2 + n_5^2 + n^2 +{\bar n}^2)  \nonumber \quad . 
\ea
More details on the explicit solution of this model may be found in ref. \cite{pss3}.

\vskip 24pt
\section{The M\"obius-Strip Amplitude}

\vskip 15pt
\input epsf \centerline{ \epsfbox{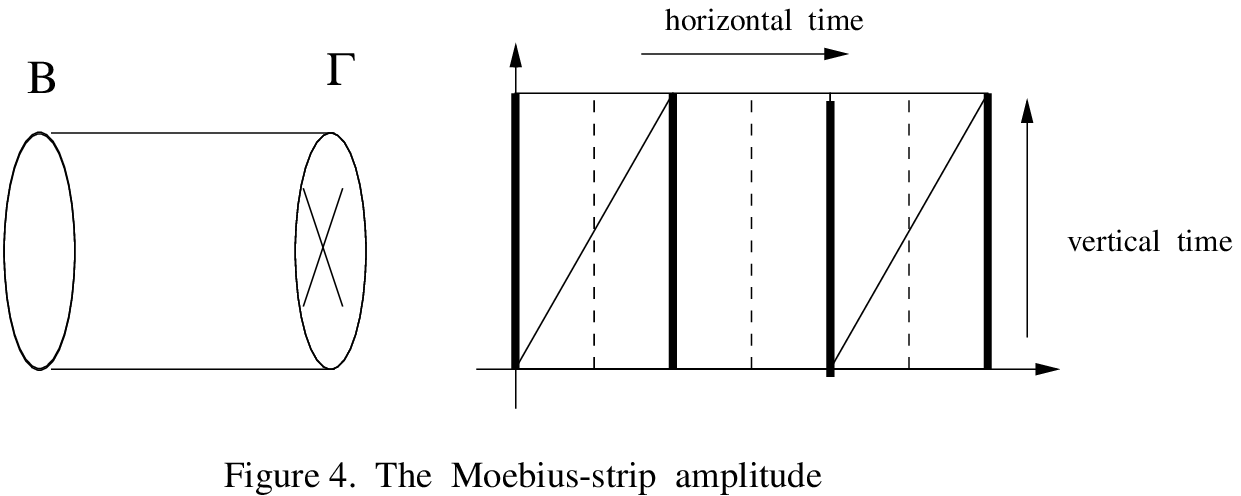}}
\vskip 15pt

The M\"obius-strip amplitude in fig. 4 is the last surface in our construction of the
spectrum. Though
perhaps more familiar than the Klein bottle, this surface has the peculiar
feature of having a double cover with a non-vanishing real part
of $\tau$.  Its value, $1/2$, induced by the relative orientation of the
horizontal sides, has an important effect: the contributions
of a given
Verma module have in this case alternating signs that depend on the level. 
The M\"obius strip can be viewed as a tube terminating at one boundary and one
crosscap.  The crosscap may be associated with the dashed line in fig. 4, whose points are all
identified in pairs in the double cover. On the other hand, the boundary comprises
the two vertical sides of the rectangle. Even in this case there are two canonical but
inequivalent choices of time.  ``Vertical'' time exhibits the propagation of open
strings undergoing a ``flip'' in their orientation, while ``horizontal'' time exhibits
the propagation of closed strings between a boundary and a crosscap. The amplitude
corresponding to the latter choice, determined by the boundary and crosscap reflection
coefficients 
$B_{i a}$ and $\Gamma_i$, is thus
\be
\tilde M \ = \  \sum_i \ {\hat \chi}_i \ \Gamma_i \ \left[ \sum_a \ B_{i a}
\ n^a
\right]
\quad .
\label{mobiust}
\ee
In order to give a meaning to this equation, one has to make corresponding
choices for the reflection coefficients 
$B_{i a}$ and $\Gamma_i$. These are determined comparing
$\tilde{M}$ (after a $P$ transformation, to be defined shortly) to the direct
channel amplitude 
\be
M \ = \  \sum_i \ {\hat \chi}_i \ M^i_a \ n^a  \quad ,
\label{mobiusd}
\ee
and thus by consistency with the direct-channel annulus amplitude of eq.
(\ref{annulusd}). The coefficients $M^i_a$ can be interpreted as twists of the
open-string spectrum, and must be integers satisfying
\be
M^i_a \ = \ A^i_{a a} \qquad  {\rm ( mod \quad 2 )} \quad ,
\label{mod2}
\ee 
a relation slightly more general than the one between
direct Klein bottle and torus coefficients, since boundary 
operators can have multiplicities $A^i_{a a}$ greater than one.
Imposing eqs. (\ref{mod2}) singles out pairs of compatible
annulus and Klein bottle projections and completes the construction of the
open descendants.

Let us return to the relation between the two forms of the
M\"obius amplitude. As we have seen, for the 
direct channel the natural modular parameter is $(i \tau +1)/2$. On the other
hand, for the vacuum channel it is 
$(i + \tau) / 2 \tau$. Due to the non-vanishing real part of $\tau$, it is 
 convenient to work in the basis of  real characters 

\be
{\hat \chi}_j  \ = \ e^{- i \pi (\Delta_j - c/24)} \ 
\chi_j \left({i \tau +1 \over 2} \right) \quad .
\label{chihat}
\ee
The modular transformation that links
direct and transverse channels is then given by the  matrix 
\be
P \ = \ T^{1/2}\ S \ T^2 \ S \ T^{1/2}
\label{pmat}
\ee 
that, just like $S$, satisfies $P^2 = C$.

By a relation reminiscent of 
eq. (\ref{verlinde}), from $S$ and $P$ one can construct the
integer-valued matrices \cite{pss} 
\be
\left( Y_i \right )_j^k \ = \ \sum_{\ell} \ { S_{i \ell} \ P_{j \ell} \ P_{k
\ell}^{\dagger} 
\over S_{1 \ell}}  \quad , \label{y1}
\ee
that form an abelian algebra and satisfy the relations 
\be
\left( Y_i \right) . \left( Y_j \right) = \sum_{\ell} {N_{i j}}^{\ell}
\left( Y_{\ell} \right) \quad . \label{y2}
\ee
One can then write compact expressions for the M\"obius
partition function  (\ref{mobiusd}) in terms of $Y_{i j k}$. In particular, if the
torus modular invariant is built with the conjugation matrix $C$, the
boundary and (chiral) bulk spectra coincide and, as we have seen, the annulus
multiplicities are equal to the fusion rule  coefficients $N_{i j k}$. One solution for
the twists in  eq. (\ref{mobiusd}) is then $M_{i j} = Y_{j i 1}$, while the
direct-channel Klein-bottle  coefficients are in this case  $K^i = Y_{i 1 1}$. Similar
representations hold for other descendants of the same modular invariant. 
The
situation becomes more intricate whenever the boundary algebra is substantially
different from the chiral one. An important instance of this phenomenon, the
non-diagonal $D_{odd}$
$su(2)$ WZW models, suggest that one can always embed the open spectrum in an
auxiliary diagonal model.

In the complete partition function of the open sector, ${ 1 \over 2 } (A + M)$,
the signs of the $n_a$ terms in $M$ determine
the symmetry of the states under the interchange of the two (equal) Chan-Paton 
charges at the ends.
Note that one can change the overall sign of the 
M\"obius amplitude by replacing all $n_a$ with $- n_a$. 
In open-string models the signs, as well as the sizes 
of the Chan-Paton groups, are further constrained by tadpole 
conditions, while the functional measure introduces an additional factor of
two in $\tilde{M}$.

Returning to our examples, 
the M\"obius amplitudes for the $A_3$ model are standard and can be directly 
expressed in terms of $Y_{ijk}$ 
\ba
M_r^{A_3} \ &=& \  {\hat \chi}_1 (n_1 - n_2 + n_3) + {\hat \chi}_3 (n_2) 
\quad,\\
M_c^{A_3} \ &=& \  {\hat \chi}_1 (n_2) + {\hat \chi}_3 (n +{\bar n} + n_2)
 \quad .
\ea
On the other hand, the $D_5$ model exhibits non-trivial solutions of eq. 
(\ref{mod2}), since the $2 n_5^2$ in the annulus amplitudes of eqs. 
(\ref{ard5}) and ( \ref{acd5}) 
can correspond either to a $2 n_5$ or to a zero M\"obius contribution.
In the first case the two sets of states have the same twist, 
while in the second case
their twists are opposite. The expressions for the two types of descendants 
are 
\ba
M_r^{D_5} &=&  \  {\hat \chi}_1 ( n_1 - n_2 + n_3 +
n_4 - n_5 ) \ + \ {\hat \chi}_3 ( - n_1 + 2 n_5 ) \ +
\nonumber \\
& & \qquad {\hat \chi}_5 (n_1 + n_3 + n_4 ) \ + 
\ {\hat \chi}_7 ( n_1 + n_2 + n_5 )  \quad ,
\ea
and
\ba
M_c^{D_5} &=&  \  {\hat \chi}_1 (- n_1 + n_2 + n_5 ) \ + \ 
{\hat \chi}_3 ( n_1 + n + {\bar n}) \ +
\nonumber \\
& & \qquad {\hat \chi}_5 (n_1 + 2 n_5 ) \ + 
\ {\hat \chi}_7 ( n_1 + n_2 + n + {\bar n} + n_5 )  \quad .
\ea

\vskip 24pt
\begin{flushleft}
{\large \bf Acknowledgments}
\end{flushleft}

We are grateful to the Organizers of the CERN Meeting on STU Dualities and
to the Organizers of the Rome Triangle Meeting for their kind invitations.
We are also grateful to Massimo Bianchi and Gianfranco Pradisi for an 
enjoyable collaboration on several aspects of this problem.  This work 
was supported in part by E.E.C. Grant CHRX-CT93-0340.

\vfill\eject

\end{document}